\pdfoutput=1
\documentclass[prl,twocolumn, amsmath,amssymb,superscriptaddress]{revtex4-1}
\usepackage{graphicx}
\usepackage{verbatim}
\usepackage{color}
\usepackage{mathrsfs} 
\usepackage{wasysym}
\usepackage{epstopdf}

\newcommand{\ket}[1]{|#1\rangle}
\newcommand{\eqn}[1]{Eq.~(\ref{#1})}
\newcommand{\fig}[1]{Fig.~\ref{#1}}

\newcommand{\erfc}{\text{erfc} }
\newcommand{\re}{\text{Re} }

\DeclareSymbolFont{extraup}{U}{zavm}{m}{n}
\DeclareMathSymbol{\vardiamond}{\mathalpha}{extraup}{87}

\begin{document}

%\preprint{}

\title{
Power Dependent Lineshape Corrections for Quantitative Spectroscopy} %Title of paper



\author{Thomas M. Stace}
%\email[]{}
%\thanks{}
 \affiliation{ARC Centre for Engineered Quantum Systems, University of Queensland, St Lucia, Brisbane 4072}
 %\affiliation{Frequency Standards and Metrology Research Group, School of Physics, The University of Western Australia}

\author{Gar-Wing Truong}
%\email[]{Gar-Wing.Truong@physics.uwa.edu.au}
%\thanks{}
%\affiliation{Frequency Standards and Metrology Research Group, The School of Physics, The University of Western Australia}

\author{James Anstie}
%\thanks{}
 \affiliation{Frequency Standards and Metrology Group, School of Physics, The University of Western Australia}

\author{Eric F. May}
%\email[]{}
%\thanks{}
%\altaffiliation{Centre for Energy, School of Mechanical and Chemical Engineering, The University of Western Australia}}
\affiliation{Centre for Energy, School of Mechanical and Chemical Engineering, The University of Western Australia}

\author{Andr\'{e} N. Luiten}
%\email[]{}
%\thanks{}

 \affiliation{Frequency Standards and Metrology Group, School of Physics, The University of Western Australia}


\date{\today}

\begin{abstract}
The Voigt profile -- a convolution of a Gaussian and a Lorentzian -- accurately describes the absorption lines of atomic and molecular gases at low probe powers.  Fitting such to experimental spectra yields both the Lorentzian natural linewidth and the Gaussian Doppler broadening.  However, as the probe power increases saturation effects introduce spurious power dependence into the fitted Doppler width. Using a simple atomic model, we calculate power-dependent corrections to the Voigt profile, which are parametrized  by the Gaussian Doppler width, the Lorentzian natural linewidth, and the optical depth.  We show  numerically and experimentally that  including the correction term substantially reduces the spurious power dependence in the fitted Gaussian width. 
\end{abstract}


\maketitle %\maketitle must follow title, authors, abstract and \pacs

 Vapour cell spectroscopy is important for determining the properties of atomic or molecular transitions \cite{sherlock2009,harris2006}, which are well described by the  Voigt profile, the workhorse of transmission spectroscopy.  
Experimental progress has reached the point that  intensity dependent corrections to the  Voigt profile are becoming significant. For example, molecular fingerprinting and other trace gas measurement applications \cite{thorpe2008cavity,adler2010cavity} using direct frequency comb spectroscopy (DFCS) \cite{diddams2007molecular} require a quantitative link between absorption depths and gas concentrations, particularly if the probe laser intensity levels are increased to improve signal-to-noise in challenging environments. While absorption sensitivities measured using DFCS can  be determined with quantum-noise limited precision \cite{PhysRevLett.107.233002}, the  effects of optical pumping on the observed depths and line-shapes remain to be quantified. %, especially for monitoring applications involving hazardous gases in which knowledge of the absolute concentrations can be critical. 
 Similarly, optical pumping can limit accurate determinations of the Boltzmann constant, $k_B$, from measurements of Doppler-broadened line shapes \cite{truong2011,PhysRevLett.100.200801,PhysRevLett.98.250801,CODATA,borde,djerroud2009,castrillo2009}.  To optimise the trade-off between signal-to-noise limitations at very low powers,  and systematic optical pumping effects at higher powers, an accurate theory of the line shape dependence on probe intensity is needed.

In the limit of  low probe powers, where atomic populations are hardly perturbed from their thermodynamic equilibrium values,   transmission spectra are well described by an exponentiated Voigt function \cite{demtroeder1981,borde2009},
$%\begin{equation}
\mathcal{T}_\mathrm{lin}(z)= e^{-\alpha z V_\nu(\Delta)}$,
%\label{eqn:linear}
%\end{equation}
where $\alpha$ is the absorptivity, $z$ is the length of the vapour cell, $\Delta$ is the detuning from resonance (note: all times and frequencies are expressed in units of $\Gamma=\Gamma_{\textrm{natural}}/2$), $\nu=\gamma/\Gamma$ is the non-dimensional Doppler width, and $V$ is a convolution of a Gaussian and a Lorentzian:
\begin{eqnarray}
V_\nu(\Delta)&\equiv&\frac{1}{\pi^{3/2}\nu}\int_{-\infty}^\infty  \frac{e^{-(x-\Delta)^2/\nu^2}}{1+x^2} \,dx,\nonumber\\
&=&\re\{ {e^{-{(i+\Delta )^2}/{\nu ^2}} \erfc\left(-{i (i+\Delta )}/{\nu }\right)}/{(\sqrt{\pi } \nu )}\},\nonumber
\end{eqnarray}
where $\erfc(z)=2\int_z^\infty e^{-t^2} dt/\sqrt{\pi}$.% is the complementary error function.

 At low probe powers,  $\mathcal{T}_\mathrm{lin}$ describes the measured transmission spectrum very accurately; the Gaussian component arises from the Doppler shifts due to the atomic Maxwell-Boltzmann velocity distribution, while the Lorentzian relates to  atomic relaxation processes.  Fitting $\mathcal{T}_\mathrm{lin}$  to experimental measurements yields $\gamma$,  $\Gamma$ and the optical depth.   However, as the probe power increases,  perturbations to the equilibrium atomic population distribution become significant and $\mathcal{T}_\mathrm{lin}$ fails to accurately represent the transmission spectrum. In the limit of  large probe powers, atomic populations depart from  thermodynamic equilibrium: in two level atoms the ground and excited states tend to equalise; in  three level systems  population may be transferred to an optically inactive state. In either case, fitting $\mathcal{T}_\mathrm{lin}$ to the measured spectrum yields incorrect values for $\gamma$ and $\Gamma$, which acquire a spurious intensity dependence. 

In this letter, we derive the intensity dependent corrections to $\mathcal{T}_\mathrm{lin}$, which can be computed perturbatively in powers of the laser intensity.   We then show, both theoretically and experimentally, that fitting to the corrected form yields Doppler widths that are independent of intensity as they should be.  This result will enable accurate measurements of Doppler broadening in quantitative spectroscopy at much higher probe intensities, where saturation effects are not negligible.  This in turn  greatly enhances signal-to-noise in a given integration time.

We begin by computing the spectral dependence on the atomic populations for a three level atom, consisting of two ground states: one optically active, $\ket{1}$, and the other optically inactive, $\ket{2}$, and an excited state $\ket{3}$, as discussed in \cite{stace2010}.  Transitions between states $\ket{1}$ and $\ket{3}$ are optically driven, and state $\ket{3}$ can relax to either of the ground states.  %After adiabatically eliminating the coherences between the atomic states,
 The population rate equations are 
\begin{equation}
\frac{d\vec{P}}{dt}=M\cdot\vec{P},\quad M=\left(\begin{array}{ccc}\frac{\Omega^2/2}{1+\Delta ^2} & 0 & 2\beta+\frac{\Omega^2/2}{1+\Delta ^2} \\
0 & 0 &2(1-\beta)\\
-\frac{\Omega^2/2}{1+\Delta ^2}& 0 & -2-\frac{\Omega^2/2}{1+\Delta ^2}\end{array}\right),\label{eqn:rate}
\end{equation}
where $\vec{P}=\{P_1,P_2,P_3\}$, 
  $\Omega$ is the atomic Rabi frequency (proportional to the electric field amplitude), $\beta$ is the branching ratio from state $\ket{3}$ to $\ket{1}$ and $\Delta$ is the detuning between the laser frequency and the atomic transition. 

%\begin{comment}
\begin{figure}[t]
\begin{center}
\includegraphics[width=\columnwidth]{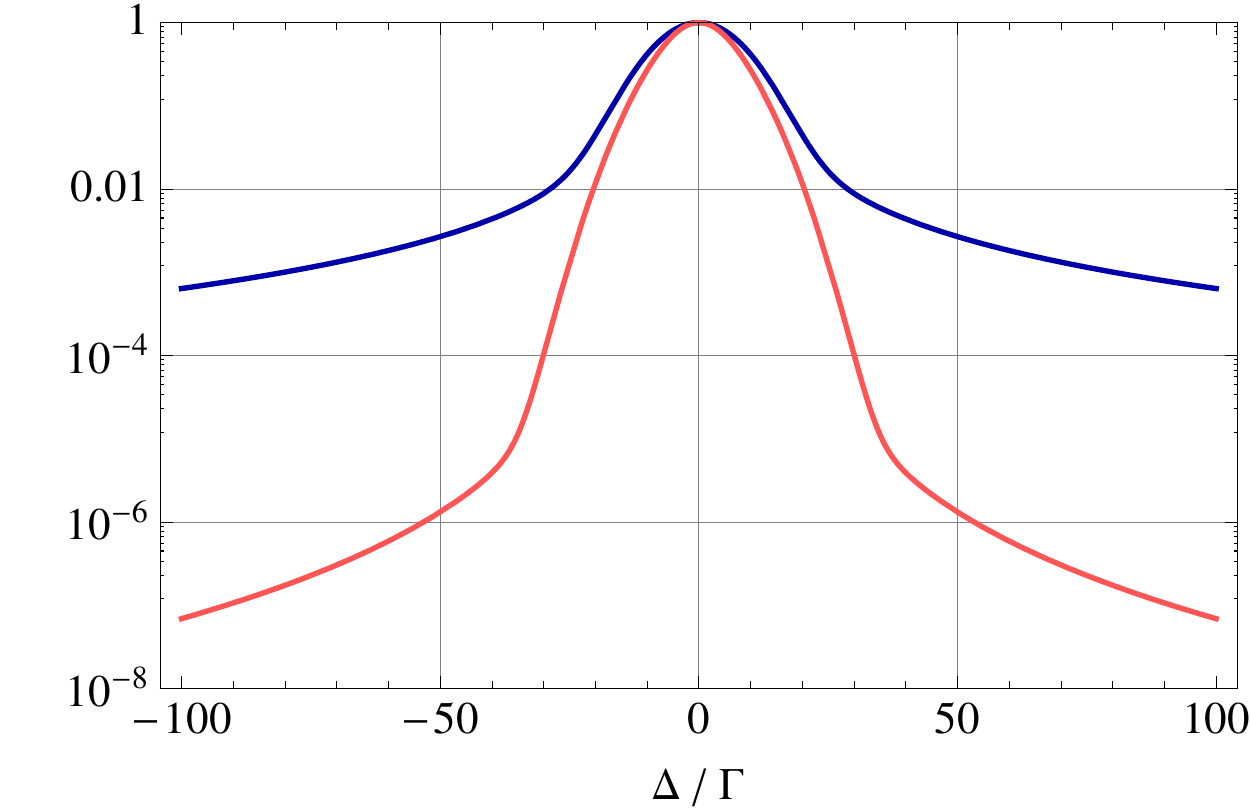}
\caption{(Colour online)   Typical form for the Voigt function (dark, blue), and the  correction term multiplying $\tilde q$ in \eqn{eqn:Acorrection} (light, red), for $\nu=10, \tilde p=20$. %and $\tilde p=-50$.
} \label{fig:CorrectionForm}
\end{center}
\end{figure}
%\end{comment}

The light is absorbed as it propagates through the atomic medium, and the axial evolution of the field is governed by \cite{stace2010}
\begin{equation}
\frac{\partial\,\ln\Omega^2}{\partial z}
=2\alpha\int_{-\infty}^{\infty} d\Delta_{v_z} \frac{e^{-{(\Delta-\Delta_{v_z})^2}/{\nu^2}}}{1+{{\Delta_{v_z}^2}}}P(\Delta_{v_z}).\label{eqn:integral}
\end{equation}
Here $P(\Delta_{v_z})=P_3(\Delta_{v_z})-P_1(\Delta_{v_z})$ and the notation $\Delta_{v_z}$ indicates that a given atoms sees a detuning which depends on its axial velocity via the Doppler shift.  If the field intensity is weak, so that on resonance populations are negligibly perturbed from thermal equilibrium, then $P=-1/2$, and the integral yields $\Omega^2(z)=\Omega^2_0 e^{\alpha z V_\nu(\Delta)}$.  Since $\mathcal{T}(z)=\Omega^2(z)/\Omega^2_0$, we recover $\mathcal{T}=\mathcal{T}_\mathrm{lin}$.

As $\Omega^2$ grows, perturbations to $P$ become significant, and we should compute corrections to $P$ using \eqn{eqn:rate}.  In the simplest case where $\Omega^2$ is time-independent \eqn{eqn:rate} can be solved analytically.  The expression is cumbersome, however it may be straightforwardly expanded in powers of $\Omega^2$ to yield
\begin{equation}
P(\Delta_{v_z})=-\frac{1}{2}+q(t)\frac{\Omega^2}{1+\Delta_{v_z}^2}+O(\Omega^4),\label{eqn:sol}
\end{equation}
where $q(t)\approx(1+\beta+2\,t(1-\beta))/8$, and $t$ is the time for which the atom is exposed to the field.  Substituting this result into \eqn{eqn:integral} yields 
\begin{equation}
\frac{\partial\,\ln\Omega^2}{\partial z}
=-\alpha\, V_\nu(\Delta)+\alpha\,q(t)\,V^{(2)}_\nu(\Delta)\,\Omega^2+O(\Omega^4)\label{eqn:integral2}
\end{equation}
where
\begin{equation}
V^{(n)}_\nu(\Delta)=\int_{-\infty}^{\infty} d\Delta_{v_z} \frac{e^{-{(\Delta-\Delta_{v_z})^2}/{\nu^2}}}{(1+{{\Delta_{v_z}^2}})^n}\nonumber%\label{eqn:V2}
\end{equation}
is a generalisation of the Voigt profile.

$V^{(2)}$ may be evaluated by noting that 
\begin{equation}
\frac{1}{(1+x^2)^2}=\lim_{a\rightarrow1}\left\{\frac{a^2}{\left(a^2-1\right) \left(a^2
   x^2+1\right)}+\frac{1}{\left(1-a^2\right) \left(x^2+1\right)}\right\}.\nonumber
\end{equation}
Convolving this   sum of two Lorentzians with a Gaussian thus yields a  sum of Voigt functions,
\begin{eqnarray}
V^{(2)}_\nu(\Delta)&=&%\lim_{a\rightarrow1}\left\{\frac{a^2}{\left(a^2-1\right) \left(1+a^2
  % x^2\right)}+\frac{1}{\left(1-a^2\right) \left(1+x^2\right)}\right\}\nonumber\\
   \lim_{a\rightarrow1}\left\{\frac{a^2}{\left(a^2-1\right) }V_{a\,\nu}(a\, \Delta)+\frac{1}{\left(1-a^2\right)}V_{\nu}( \Delta)\right\}\nonumber\\
   &=&\textrm{Re}\{\frac{ \nu +\sqrt{\pi } e^{-\frac{(\Delta +i)^2}{\nu ^2}} \left( i \Delta +\nu
   ^2/2-1\right) \erfc(\frac{1-i \Delta }{\nu })}{ \pi  \nu ^3}\}.\nonumber
\end{eqnarray}
%We note in passing that $V^{(2)}$ is a linear combination of $V$ and its second derivative. 
Higher order correction terms can be calculated iteratively, using the same method.

%\begin{comment}
 \begin{figure}
\begin{center}
\includegraphics[width=\columnwidth]{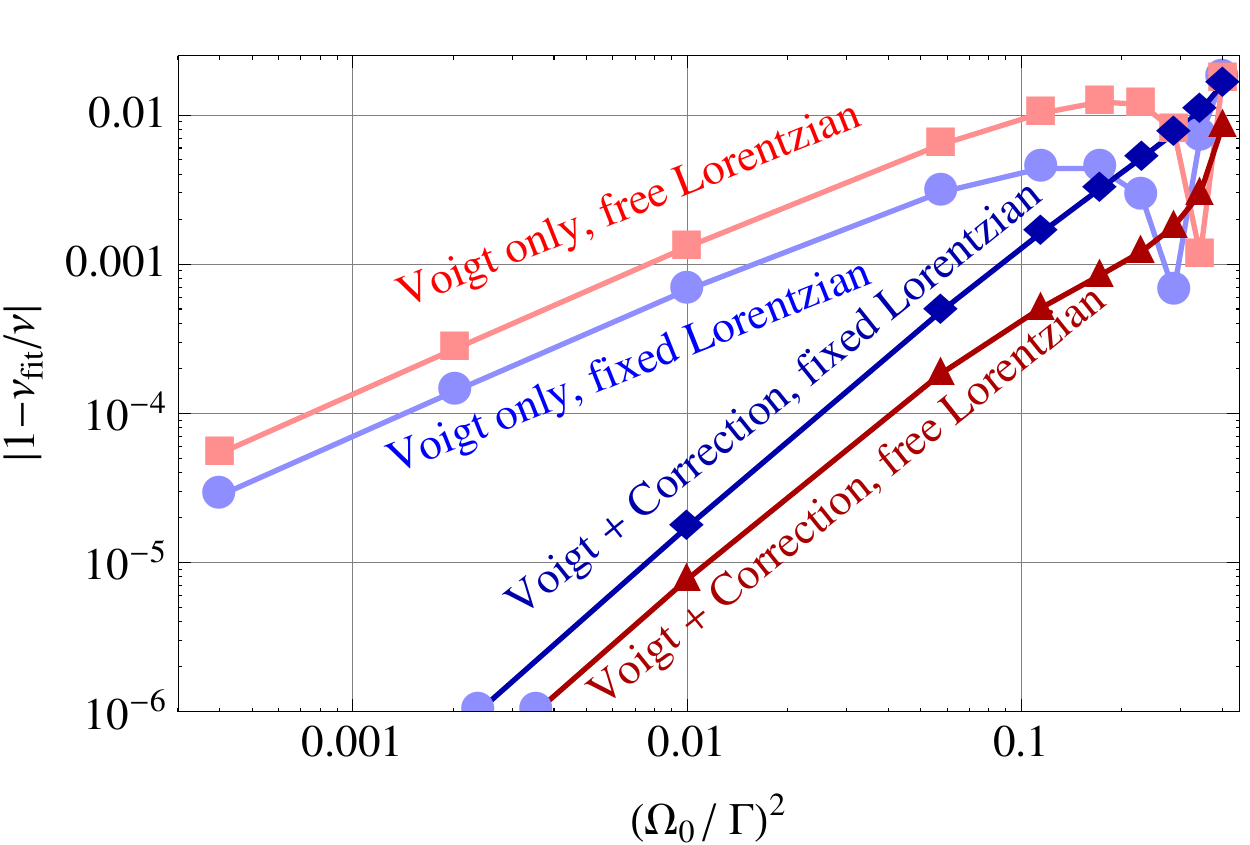}%\includegraphics[width=\columnwidth]{FittednuLogexp}
\caption{(Colour online) Fits to numerically simulated spectra, showing the ratio of fitted Doppler to lorentz widths demonstrate the improved performance of fits using the correction term.} \label{fig:lognu}
\end{center}
\end{figure}
%\end{comment}

%\begin{comment}
\begin{figure*}
\begin{center}

\includegraphics[width=\columnwidth]{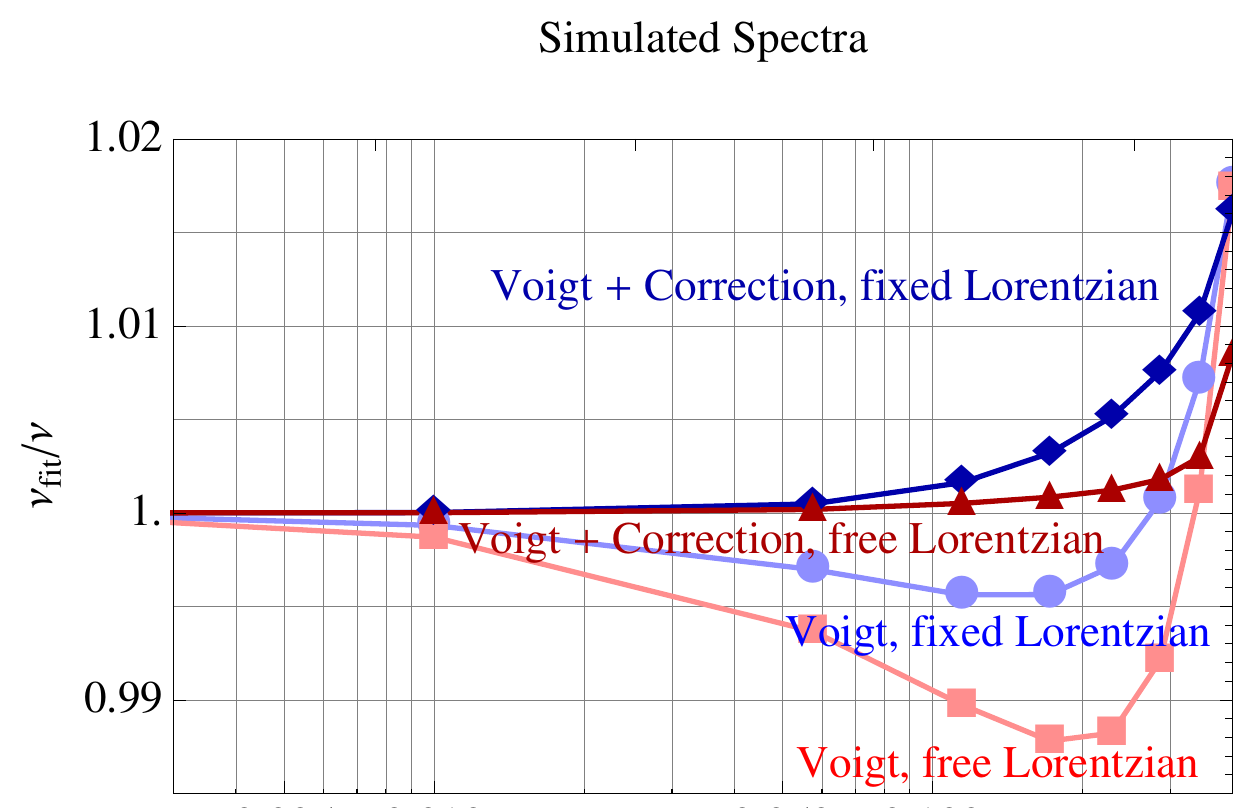}
\includegraphics[width=\columnwidth]{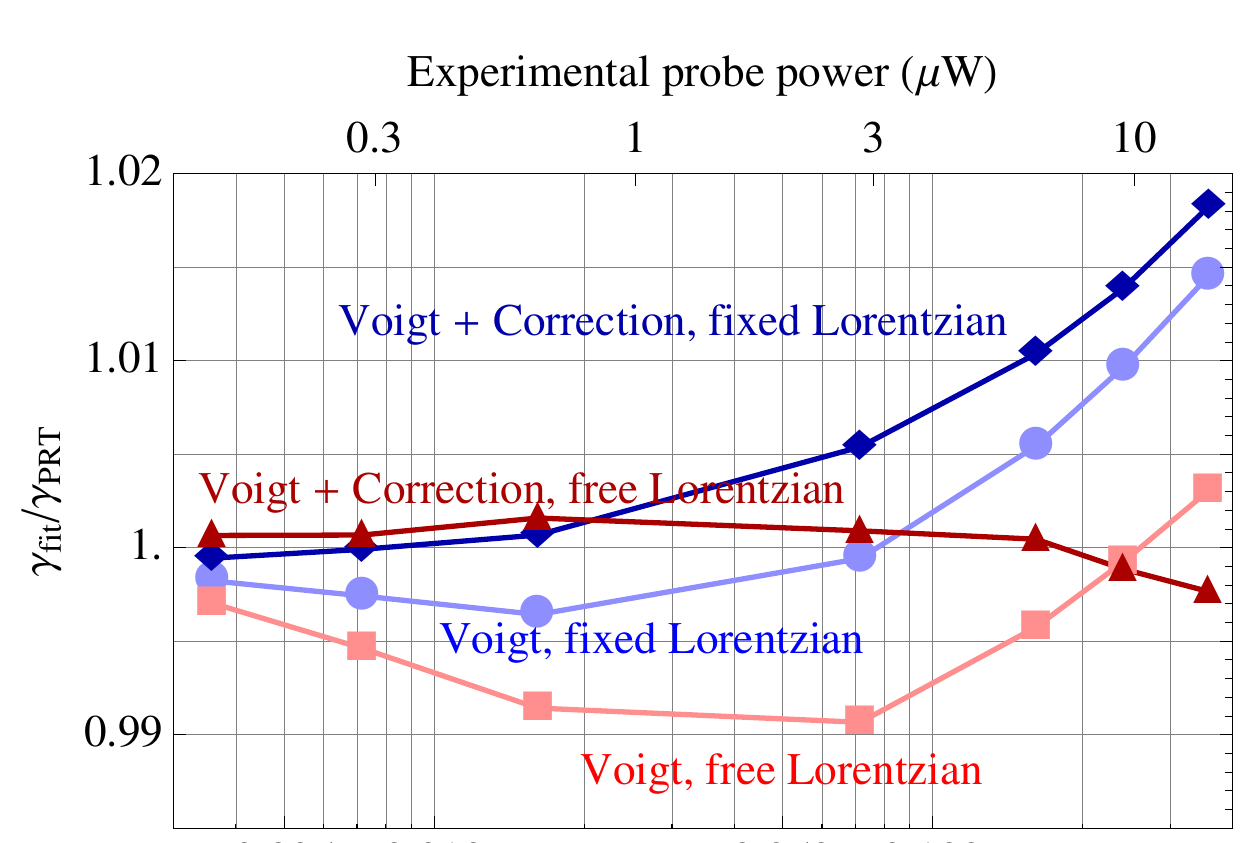}
\includegraphics[width=\columnwidth]{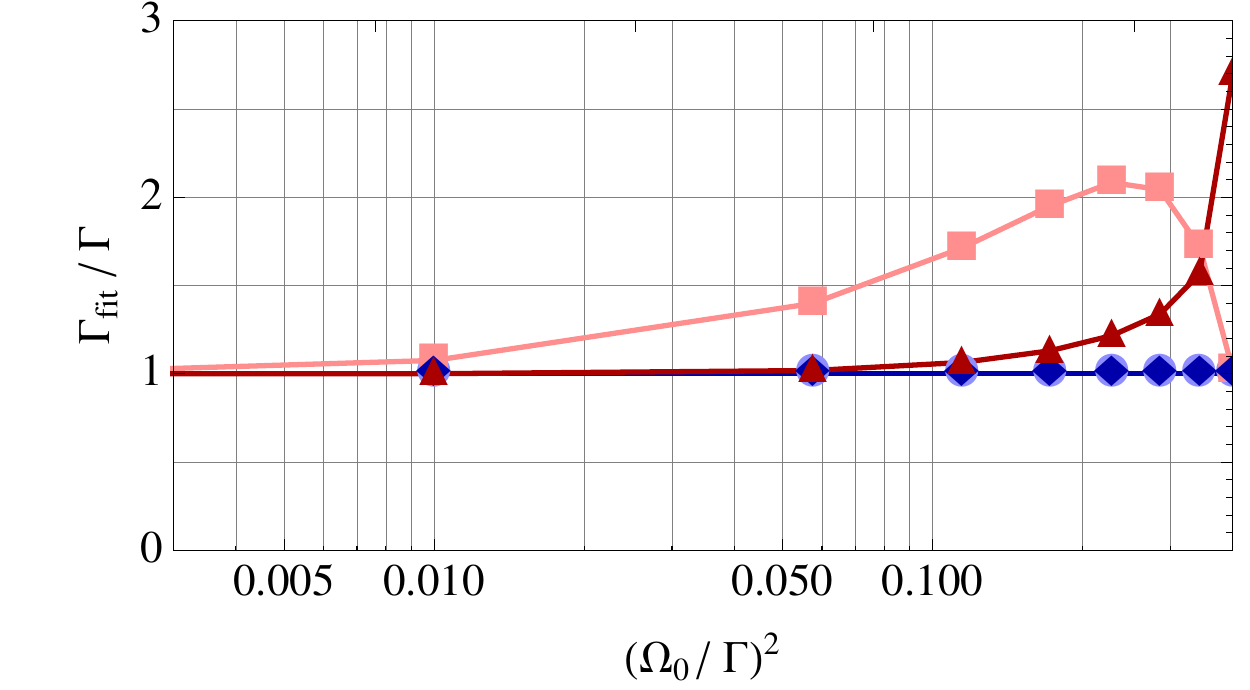}
\includegraphics[width=\columnwidth]{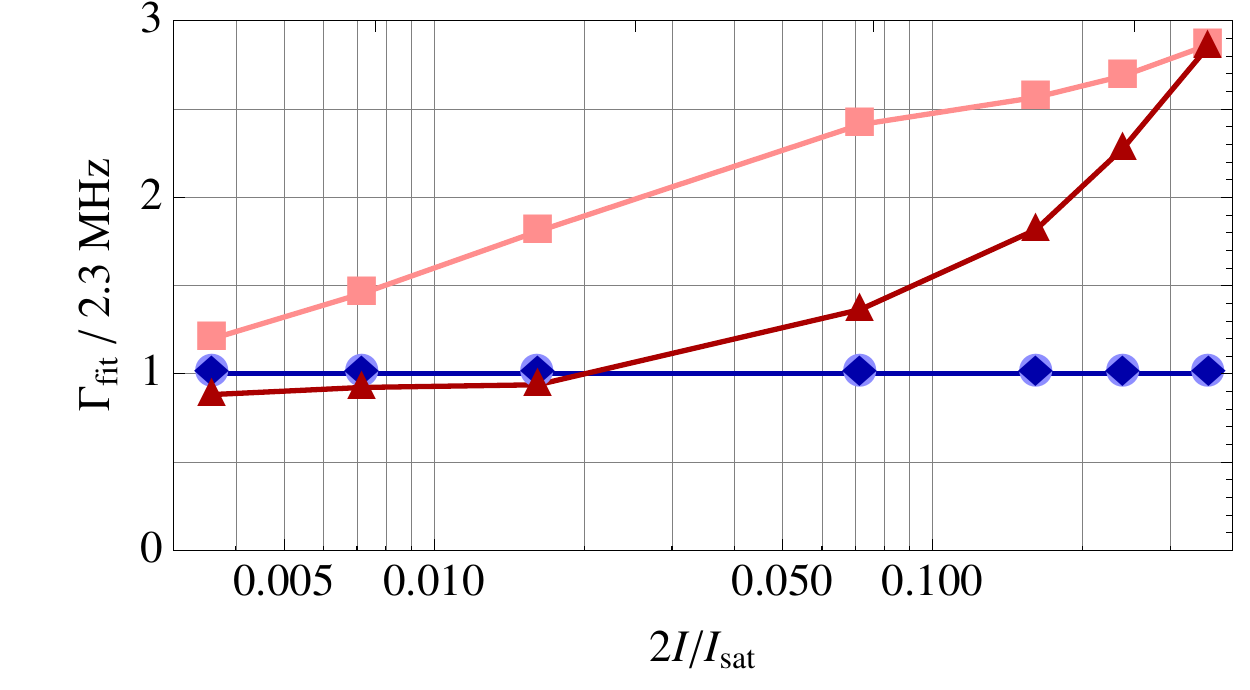}

\caption{(Colour online)  Fits to (left) Simulated data using \eqn{eqn:integral}, with $\beta=1/2$, $t=20$, $\nu=100$; (right) experimental data taken from  \cite{truong12}. (top) Fitted Doppler widths,  extracted from fits to Voigt only or Voigt plus correction, with fixed or free Lorentzian widths. (bottom) Fitted Lorentzian widths, normalised to the natural linewidth.} \label{fig:simulatedpowerdependence}
\label{fig:experimentalpowerdependence}
\end{center}
\end{figure*}
%\end{comment}

Equation (\ref{eqn:integral2}) can be solved analytically to give
\begin{eqnarray}
\mathcal{T}&=&\frac{    \,e^{-\alpha  z V_{\nu }(\Delta )} }{1+q\,
   \Omega _0^2 \,V^{(2)}_{\nu }(\Delta ) (e^{-\alpha  z V_{\nu }(\Delta )}-1)/  V_{\nu }(\Delta )},\label{eqn:central}
\end{eqnarray}
Expanding this expression, and recalling that all quantities are nondimensionalised in units of $\Gamma$ we explicitly include the Lorentz width as a parameter
\begin{eqnarray}
\mathcal{T}
    &\approx& e^{{\tilde p V_{\nu }(\Delta/\Gamma )+\tilde q
    (e^{\tilde p
   V_{\nu }(\Delta/\Gamma )}-1)\,V^{(2)}_{\nu }(\Delta/\Gamma )/  V_{\nu }(\Delta /\Gamma)}}
   %+O(\Omega_0^4)
   \label{eqn:Acorrection}
\end{eqnarray}
where  $\tilde p=-\alpha\,z$ and $\tilde q= -q\,\Omega_0^2$.  
This correction, linear in $\tilde q\propto I$, is the central result of this Letter.  If  either the intensity ($\propto\tilde q$) is small, or if the optical depth ($\propto \tilde p$) is small, the correction vanishes and the well-known Voigt profile is recovered.  In the latter case, saturation effects may be substantial, however the optical depth of the sample is sufficiently short that only a small fraction of incident photons are absorbed. 
    In what follows we demonstrate both numerically and experimentally the advantages of using   \eqn{eqn:Acorrection} when fitting spectra  for which saturation effects become significant.

We note that the derivation of the correction term assumed the Rabi frequency was time independent.  However, in an experiment atoms crossing through a probe laser beam see a time-dependent field.  Nevertheless, \eqn{eqn:sol} and (\ref{eqn:integral2}) are still correct, albeit with some more complicated time dependence in $q(t)$, and  taking $\Omega$ to represent a typical scale of the local Rabi frequency (e.g.\ that corresponding to the peak intensity in the probe).    It follows that the form of the correction term in \eqn{eqn:Acorrection} is also valid, where the complicated time dependence in $q$ is simply absorbed into the parameter $\tilde q$.

 A typical Voigt profile and the correction are shown in \fig{fig:CorrectionForm}. In fitting to spectra,  $\nu$, $\tilde p$ and $\tilde q$ are treated as fitting parameters.  We also numerically investigate leaving $\Gamma$ free, which heuristically accounts for higher order intensity dependences that we have otherwise neglected; variation in $\Gamma$  implies a breakdown in the fitting model.

To evaluate the efficacy of the correction, we  numerically simulate an absorption spectrum using  \eqn{eqn:integral} (keeping terms in $P$ up to $O(\Omega^6)$), and then attempt to extract the simulation parameters by fitting \eqn{eqn:Acorrection}. For comparison, we compare fits with Voigt only ($\tilde q=0$) and  Voigt plus correction ($\tilde q$ free); in combination with a fixed Lorentzian ($\Gamma=0$) and free Lorentzian ($\Gamma$ free). \fig{fig:lognu} shows the deviation of $\nu_{\textrm{fit}}=\gamma_{\textrm{fit}}/\Gamma$ from the correct value, $\nu$, as a function of  intensity ($\propto(\Omega_0/\Gamma)^2$), for the  different fitting forms.  As expected, all fitting forms yield the correct value for $\nu_{\textrm{fit}}$ as $\Omega_0^2\rightarrow 0$.  Without the correction ($\blacksquare$ and $\CIRCLE$), the fit converges to the correct value as $\Omega_0^2$. This improves to $\Omega_0^4$ when the correction is included ($\vardiamond$ and $\blacktriangle$).  

The fitting performance is further illustrated in \mbox{\fig{fig:simulatedpowerdependence} (left)}, which  shows the dependence of the fitted parameters $\nu_{\textrm{fit}}/\nu$ (top) and $\Gamma$ (bottom) as a function of the probe intensity ($\propto(\Omega_0/\Gamma)^2$, for a particular representative choice of  parameters ($\beta=1/2$, $t=20$, $\nu=100$).   Also shown is the fitted Lorentzian component, \fig{fig:simulatedpowerdependence} (bottom), demonstrating apparent  broadening of the  atomic lifetime, even for intensities substantially below the saturation intensity.  This phenomenon has been observed experimentally \cite{truong12}.

Qualitatively, at low powers the uncorrected, Voigt-only fits ($\blacksquare$ and $\CIRCLE$) systematically underestimate $\nu$ over a range of laser powers, and demonstrate a pronounced minimum, as illustrated in \fig{fig:simulatedpowerdependence}(top).  In the case where  the Lorentzian is held fixed ($\CIRCLE$),  the Voigt-only fit may over- or under-estimate the Doppler width, depending on the parameters, i.e.\ the initial slope may be positive or negative.
 For the parameter choices in the simulated spectra, \mbox{\fig{fig:simulatedpowerdependence} (left)}, the initial slope is slightly negative, and there is a noticeable minimum in $\nu_{\textrm{fit}}/\nu$.

We now demonstrate the effectiveness of using the corrected transmission function for extracting the true atomic velocity distribution from experimental data.  Spectra were recorded in a 7 cm long Caesium vapour cell at approximately 295 K at different probe powers, using a collimated  2 mm diameter laser. At each power setting, 16 spectral scans were taken, and each scan was fitted using \eqn{eqn:Acorrection} to yields estimates for $\gamma_{\textrm{fit}}$ and $\Gamma_{\textrm{fit}}$.    For each scan, the temperature was recorded using a platinum resistance thermometer (PRT), accurate to $\pm100$ ppm.  Each fitted Doppler width, $\gamma_{\textrm{fit}}$, was normalised by the width expected given the temperature measured using the PRT,  \mbox{$\gamma_{\textrm{PRT}}=\omega_{\textrm{probe}}\sqrt{2k_BT_{\textrm{PRT}}/m}/c$}. 

To qualitatively compare the fitting results from the simulated spectra  with the experimental results, we convert the laser power to an average intensity by dividing the power by the beam area, $\pi (1~\textrm{mm})^2$, and then normalise this average intensity by the saturation intensity $I_{\textrm{sat}}= 2.51~\textrm{mW/cm}^2$ \cite{steck2003cdl}, so that the nondimensional  experimental parameter,  $2I/I_{\textrm{sat}}$,  is equivalent to the nondimensional simulation parameter $(\Omega/\Gamma)^2$.  

The results are shown in \fig{fig:experimentalpowerdependence} (right).  For \mbox{$2I/I_{\textrm{sat}}<0.1$}, the extracted Doppler widths are substantially better when using the correction than not, as shown in \fig{fig:experimentalpowerdependence} (right,top).  This is in qualitative agreement with the simulated data. For the lowest powers used, the corrected estimates for the Doppler width  are at least an order of magnitude better than the uncorrected form, limited  by the accuracy of the PRT calibration, and statistical uncertainty from the small sample size.  

  Comparing \fig{fig:experimentalpowerdependence} (left) and (right) shows that the simulated results are  in qualitative agreement with the experimentally derived results in other respects as well, including the relative locations of minima, slopes and crossing points. Quantitative differences arise from the simplified  physical model underpinning the simulated spectra, in which the distribution of atomic crossing times and intensity profiles is replaced by the average crossing time and the average intensity. While not unphysical, this differs from the  experimental conditions. 

In thermometry, where the objective is to extract very accurate Doppler widths from spectral data, we see from \fig{fig:lognu} that it is important to use the Voigt correction.  For instance, if $2I/I_{\textrm{sat}}\sim 0.003$ then a fit to a Voigt-only profile has a systematic bias in the fitted Doppler width at the level of several hundred ppm.  In contrast, using the  corrected form yields an estimate of the Doppler width that is accurate to $\sim 1$ ppm, which is comparable to the current thermometric state-of-the-art.

It follows that  under the comparable experimental conditions represented in \fig{fig:experimentalpowerdependence} (right), to reduce the systematic bias in the fitted Doppler width to 1 ppm (when utilising the corrected form), the probe beam should operate at $2I/I_{\textrm{sat}}= 0.003$,  corresponding to a laser power of $ 100$ nW in a 2 mm diameter beam.

One important application of Doppler spectroscopy is to contribute to the CODATA redefinition of Boltzmann's constant \cite{PhysRevLett.98.250801,djerroud2009,castrillo2009,PhysRevLett.100.200801}, in which the vapour cell is at equilibrium with a triple-point-of-water reference at $273.16$ K.  At this temperature the optical depth of the cell is $\sim 10$ times smaller than under the experimental conditions reported here.  As such, the intensity may be increased by the same factor, to  $\sim1\, \mu$W, while still maintaining the systematic Doppler error at $\sim1$ ppm.
  
In conclusion, we have derived the power-dependent correction to the Voigt profile, and demonstrated numerically and experimentally that including the correction term yields much better estimates of the width of the underlying Gaussian process, substantially reducing the spurious power dependences that arise from an uncorrected Voigt profile. We anticipate that this will find direct application in the near-term to high-precision spectroscopy and thermometry of atomic vapour cells.

\bibliography{bib}

\end{document}